\documentclass{ws-procs961x669}            
\usepackage{ws-toc,ws-multind}

\makeindex{author}                     

\begin{document}
\def\fnl{f_{_{\rm NL}}}
\def\d{{\rm d}}
\def\pp{p_{\phi}}
\def\dpp{\delta p_{\phi}}
\def\dph{\delta\phi}
\def\u{\mathfrak{A}}
\def \almx{a_{_{\ell_1 m_1}}^{^X}}
\def \almy{a_{_{\ell_2 m_2}}^{^Y}}
\def \almz{a_{_{\ell_3 m_3}}^{^Z}}
\def\Mpc{{\rm Mpc}}
\wstoc{For proceedings contributors: Using World Scientific's\\ WS-procs961x669 document class in \LaTeX2e}
{Roshna K and V. Sreenath}

\title{Viability of loop quantum cosmology at the level of bispectrum}

\author{Roshna K $^*$ And V. Sreenath}
\index{author}{Roshna, K.} 
\index{author}{Sreenath, V.} 

\address{Department of Physics, National Institute of Technology Karnataka, Surathkal,\\
Mangaluru 575025, India.\\
$^*$E-mail: roshnak.217ph005@nitk.edu.in\\
sreenath@nitk.edu.in}



\begin{abstract}
Observations by Planck indicate that CMB anisotropies are consistent with predictions of nearly Gaussian primordial perturbations as the one generated in slow-roll inflation. On the other hand, loop quantum cosmology (LQC)  generates a non-Gaussian bispectrum. In particular, calculations of primordial bispectrum generated in LQC shows that the non-Gaussianity function $f_{_{\rm NL}}(k_1,\, k_2,\, k_3)$  is highly scale-dependent and oscillatory at long wavelengths and is nearly scale-invariant as in slow-roll at small scales. We discuss the viability of such a non-Gaussian bispectrum in the light of observations by Planck. More specifically, we model the bispectrum generated in LQC and compute its imprints on the CMB bispectrum. We then show that the CMB bispectrum generated in LQC though non-Gaussian, due to its highly oscillatory nature, is similar to that generated in slow-roll inflation and hence consistent with the observations by Planck. 
\end{abstract}


\bodymatter

\section{Introduction}\label{rs:sec1}
Falsifiability is a basic tenet of Science. 
It is the idea that for any scientific hypothesis to have credence, it should be disprovable. 
This is true for Cosmology as well. 
In Cosmology, it is generally believed that it is the tiny primordial fluctuations in the fabric of spacetime, generated in the early universe, that leads to the anisotropies in the Cosmic Microwave Background (CMB) and inhomogeneities in the Large Scale Structure (LSS). Advances in techniques to observe CMB and LSS have enabled us to test and verify this idea. For, any model which provides a mechanism to generate primordial perturbations must explain the observed CMB and LSS. 
\par 
Primordial perturbations being a stochastic quantity are quantified using their correlation functions. If the perturbations are Gaussian, it can be fully described by its power spectrum. Any deviation from Gaussian nature is quantified using higher-order correlations. The strength and scale dependence of these correlations can be inferred by studying correlations of anisotropies in CMB and inhomogeneities in LSS. 
Observations of CMB \cite{WMAP:2012fli, WMAP:2012nax, Planck:2018nkj, Planck:2018vyg, Planck:2018jri, Planck:2019kim} indicates that the standard model of cosmology namely $\Lambda$CDM model together with inflation provides a robust framework for explaining the evolution and structure of our Universe. According to this model, it is the tiny quantum fluctuations produced during inflation which gave rise to primordial perturbations (see, for instance, Refs.~\citenum{Riotto:2002yw, Sriramkumar:2009kg, RevModPhys.78.537, Baumann:2009ds}). 
Observations of CMB anisotropies from the Planck \cite{Planck:2018nkj} satellite indicate that the anisotropies observed in the CMB are consistent with that caused by a primordial perturbation which is Gaussian and nearly scale-invariant as predicted by slow-roll inflation (see, for instance,
Refs.~\citenum{Planck:2018jri, Planck:2019kim, Martin:2013tda}). Despite the success of inflationary paradigm, there are some questions that need to be answered. For instance, inflation does not address the initial big bang singularity, leaving the origin and evolution of perturbations in the planck regime unexplained and hence the past incomplete.
\par 
Loop quantum cosmology (LQC) \cite{Bojowald:2001xe, PhysRevLett.96.141301, Ashtekar:2006wn, Bojowald:2008jv,   Agullo:2012sh, Agullo:2012fc, Agullo:2013ai, Agullo:2015tca, Ashtekar:2015dja, PhysRevD.97.066021} is an effort to extend the inflationary regime to the planck regime using principles of loop quantum gravity. For reviews on LQC refer to Refs.~\citenum{Ashtekar:2011ni, Agullo:2013dla, Agullo:2016tjh, Agullo:2023rqq}. In LQC, big bang singularity is replaced by a quantum bounce. In this scenario, perturbations begin their evolution in the quantum vacuum at some early time and then evolve through the bounce and the inflationary epoch, leaving their imprints on the CMB. One of the important effects of the bounce is to introduce an additional scale corresponding to the curvature at the bounce named as $k_{_{\rm LQC}}$. Studies show that, the modes which have comparable lengths to this new scale gets modified, leading to a highly scale-dependent power spectrum. Furthermore, perturbations generated in LQC are highly non-Gaussian in nature. The non-Gaussainity function $f_{_{\rm NL}}(k_1,\, k_2,\, k_3)$, is found to be highly scale-dependent and oscillatory at much longer wavelengths {\it i.e.} $k<< k_{_{\rm LQC}}$ and is nearly scale-invariant for shorter wavelengths as in the case of slow-roll inflation. It is interesting to check the viability of such a highly oscillatory and scale-dependent bispectrum generated in LQC in the light of observations by Planck. In this article, we discuss whether the scale-dependent and non-Gaussian perturbation generated in LQC is consistent at the level of bispectrum with a nearly scale-invariant and nearly Gaussian perturbations observed by Planck. 
\par 
This article is organised as follows. In the next section, we review essentials of generation of primordial perturbations in LQC. We then describe the numerical results for power spectrum and bispectrum. We argue that both spectra are consistent with observations if the pre-inflationary expansion following the bounce is large enough. In \sref{rs:sec3}, we consider the case in which the duration of pre-inflationary expansion is sufficient enough that the effect of  bounce appears at the observable scales. By studying CMB bispectrum, we show that even in this scenario, predictions of LQC will be consistent with observation. We conclude the paper with a summary and discussion of results. 
\par
Throughout this paper $\kappa\, =\, 8\,\pi\,G$, where $G$ is the Newton's constant and $H\,=\,\dot a/a$ is the Hubble parameter with $a$ being the scale factor. Further, an overdot and a prime refers to a derivative with respect to cosmic time and conformal time respectively. 

\section{Primordial perturbations generated in LQC\label{rs:sec2}}
In LQC, the dynamics of our universe is described by a wavefunction $\Psi(v, \phi, \delta\phi)$ where $v = {\cal V}_0 a^3/\kappa$ is the volume of the universe, $\phi$ is the scalar field which sources the background Friedmann-Lema\^itre-Robertson-Walker (FLRW) spacetime and $\delta \phi$ is the perturbation to the scalar field living on this background. The wavefunction satisfies an equation similar to the Wheeler-DeWitt equation of the form $\hat {\cal H}\Psi(v, \phi, \delta\phi)\, =\, 0$. In the absence of perturbations, if the wavefunction is sharply peaked \cite{PhysRevLett.96.141301, Ashtekar:2006wn, Diener:2013uka, Agullo:2016hap} over the volume, it has been shown that the quantum dynamics of the FLRW spacetime can be described using modified Friedmann and Raychaudhury equations,  {\it viz.} 
\begin{subequations}
\begin{align}\label{rs:eq1}
    H^2\, =\, \frac{\kappa}{3}\rho\left(1\,-\, \frac{\rho}{\rho_{\rm sup}} \right)\\
    \frac{\ddot a}{a}\, =\, -\frac{\kappa}{6}\,\rho\,\left( 1\, -\, 4\,\frac{\rho}{\rho_{\rm sup}} \right)\, -\, \frac{\kappa}{2}\, P\,\left( 1\,-\,2\,\frac{\rho}{\rho_{\rm sup}} \right)
\end{align}
\end{subequations}
where $\rho\, =\, \frac{\dot\phi^2}{2}\,+\,V(\phi)$ and $P\,=\, \frac{\dot\phi^2}{2}\,-\,V(\phi)$ are the energy density and pressure of the background scalar field respectively. Above equations indicate that universe undergoes a bounce at a supremum energy density of $\rho\, =\,\rho_{\rm sup}$. The scalar field is governed by the equation $\ddot\phi\,+\, 3\,H\,\dot\phi\,+\,V_\phi\,=\,0$. Studies show that, for a suitable potential, universe transitions to a slow-roll phase after the bounce (see, for instance, Refs.~\citenum{Ashtekar:2009mm, Ashtekar:2011rm, Bolliet:2017czc, PhysRevD.99.063520}). 
\par
We shall work with the {\sl dressed metric}\cite{Agullo:2012sh, Agullo:2012fc, Agullo:2013ai, PhysRevD.97.066021} approach to study perturbations in LQC. In this approach, we are interested in solutions of the form $\Psi(v, \phi, \delta\phi)\,=\,\Psi^{(0)}(v, \phi)\otimes\delta\Psi(v, \phi, \delta\phi)$, where $\Psi^{(0)}(v, \phi)$ describes the quantum FLRW spacetime and $\delta\Psi(v, \phi, \delta\phi)$ represents perturbations. Thus, in this approach, perturbations are treated as test fields living on the quantum FLRW background. This amounts to working with classical equations of perturbations with the background functions replaced by the solutions of effective equations.  

\subsection{Primordial Correlations}
Primordial perturbations being random variables are quantified using correlation functions. To compute these functions, we expand $\hat{\delta \phi}$ in Fourier space by introducing annihilation and creation operators as (see, for instance, Ref.~\citenum{PhysRevD.97.066021}),  
\begin{equation}\label{rs:eq2}
	\hat{\delta\phi}({\vec x},\eta) = \int\frac{{\rm d}^3 k}{(2\pi)^3} \left(\hat A_{\vec k}~\varphi_k(\eta) + \hat A^\dagger_{-\vec k}
	~\varphi_{k}^*(\eta)\right) e^{i{\vec k}\cdot{\vec x}}.
\end{equation}
The evolution of the Fourier mode $\varphi_{k}$ is given by
\begin{equation}\label{rs:eq3}
\varphi_k^{\prime\prime} + 2\frac{a '}{a} \varphi_k^\prime + (k^2 +
a^2 \, {\cal U})\,  \varphi_k =0\, ,
\end{equation}
where  $k^2\equiv  k_i k_j\, \delta^{ij}$ is the comoving wavenumber and ${\cal U}$ is a background dependent potential. 
The scalar power spectrum $ \mathcal P_{\delta\phi}(k, \eta)$ of $\hat \delta \phi$, which quantifies the two-point correlation at a time $\eta$, is defined as,
\begin{equation} \label{rs:eq4}
	\langle 0|\hat{\delta\phi}_{\vec k}( \eta) \hat{\delta\phi}_{\vec k^\prime}(\eta)|0\rangle \equiv 
	(2\pi)^3\delta^{{(3)}}({\vec k}+{\vec k^\prime}) \frac{2\pi^2}{k^3} \mathcal P_{\delta\phi}(k, \eta).
\end{equation}
\par 
In the presence of self-interactions, higher-order correlations can be generated. Of lowest order among them is the three-point correlation of $\hat{\delta\phi}$, which at tree level is given as (see, for instance, Refs.~\citenum{PhysRevD.97.066021, maldacena2003non}), 
\begin{eqnarray}\label{rs:eq5}
	\langle0|\, \hat{\delta\phi}_{\vec k_1}( \eta)\,\hat{\delta\phi}_{\vec k_2}( \eta)\hat{\delta\phi}_{\vec k_3}( \eta)\,|0\rangle &=&\, 
	-\,i/\hbar  \int d \eta' \langle 0|\left[ \hat{\delta\phi}^{\rm I}_{{\vec k}_1}( \eta) \hat{\delta\phi}^{\rm I}_{{\vec k}_2}(\eta) \hat{\delta\phi}^{\rm I}_{{\vec k}_3}( \eta), \hat{\mathcal H}^{\rm I}_{\rm int}( \eta')\right]|0\rangle \nonumber\\
	&&\,+\,\mathcal{O}(\mathcal H^2_{\rm int}),
\end{eqnarray}
where $\hat{\mathcal H}^{\rm I}_{\rm int}( \eta)$ is the interaction Hamiltonian computed at third order in perturbations. 
\par 
Since comoving curvature perturbation ${\mathcal R}$ remain constant after crossing the horizon, it is convenient to express correlations in terms of ${\mathcal R}$. 
It is related to perturbations in scalar field $\delta \phi$ as,
\begin{equation}\label{rs:eq6}
	 \mathcal{R}(\vec x,\eta )=- \frac{a}{z} \, \dph(\vec x,\eta)+\left[-\frac{3}{2}+3\frac{V_{\phi}\, a^5}{\kappa\, P_\phi\, \pi_a}+\frac{\kappa}{4}\frac{z^2}{a^2}\right]  \left(\frac{a}{z} \, \delta\phi(\vec x, \eta)\right)^2+\cdots \, ,
\end{equation}
where trailing dots indicate subdominant terms at higher order in perturbations.  
In terms of ${\mathcal R}$,  the power spectrum is given by
\begin{equation}\label{rs:eq7}
	{\mathcal P}_{\cal R}(k)\, =\, \biggl( \frac{a(\eta_{f})}{z(\eta_{f})}\biggr)^2\,{\mathcal P}_{\delta\phi}(k,\,\eta_{f}),
\end{equation}
where $z\,=\,-6\,p_{\phi}/(\kappa\,\pi_a)$ with $\pi_a$ and $p_{\phi}$ being the conjugate momentum to scale factor and scalar field respectively. Time $\eta_f$ indicates an instant well after the modes have crossed the horizon during inflation. 
The three-point correlation function of ${\mathcal R}$ is, 
\begin{eqnarray}\label{rs:eq8}
	& & \langle 0|\hat{\mathcal{R}}_{{\vec k}_1} \hat{\mathcal{R}}_{{\vec k}_2} \hat{\mathcal{R}}_{{\vec k}_3}|0\rangle=\left(-\frac{a}{z}\right)^3  \langle 0|\hat{{\delta\phi}}_{{\vec k}_1} \hat{{\delta\phi}}_{{\vec k}_2} \hat{{\delta\phi}}_{{\vec k}_3}|0\rangle + \left(-\frac{3}{2}+3\frac{V_{\phi}\, a^5}{\kappa\, p_\phi \, \pi_a}+\frac{{\kappa}}{4}\frac{z^2}{a^2}\right)\,  \left(-\frac{a}{z}\right)^4\, \nonumber \\ & & \times \Big[\int \frac{d^3p}{(2\pi)^3} \, \langle 0|\hat{\delta\phi}_{{\vec k}_1} \hat{{\delta\phi}}_{{\vec k}_2}  \hat{{\delta\phi}}_{{\vec p}}\,  \hat{{\delta\phi}}_{{\vec k}_3-\vec p}|0\rangle + (\vec k_1 \leftrightarrow \vec k_3)+ (\vec k_2 \leftrightarrow \vec k_3) \, 
	+\cdots\Big] \, .  
\end{eqnarray}
The observable quantity computed at $\eta_f$ is the scalar bispectrum $B_{\mathcal{R}}(k_1,k_2,k_3)$. It is defined as, 
\begin{eqnarray}\label{rs:eq9}
	\langle 0|\hat{\mathcal{R}}_{{\vec k}_1} \hat{\mathcal{R}}_{{\vec k}_2} \hat{\mathcal{R}}_{{\vec k}_3}|0\rangle\equiv  (2\pi)^3\delta^{(3)}(\vec{k}_1+\vec{k}_2+\vec{k}_3) \, B_{\mathcal{R}}(k_1,k_2,k_3) \, .
\end{eqnarray}
The amplitude of the primordial bispectrum is usually quantified by a dimensionless function  $f_{_{\rm NL}}(k_1,\,k_2,\, k_3)$ and is related to the scalar bispectrum as,
\begin{eqnarray}\label{rs:eq10}  f_{_{\rm NL}}(k_1,k_2,k_3) &\equiv&  -\frac{5}{6}
	(2\,\pi^2)^{-2} \biggl(\, \frac{{\cal P_R}(k_1)}{k_1^3}\,\frac{{\cal P_R}(k_2)}{k_2^3}\,+\, \frac{{\cal P_R}(k_2)}{k_2^3}\,\frac{{\cal P_R}(k_3)}{k_3^3}\, \nonumber\\
	&&+\, \frac{{\cal P_R}(k_3)}{k_3^3}\,\frac{{\cal P_R}(k_1)}{k_1^3}\biggr)^{-1}\, B_{\mathcal{R}}(k_1,k_2,k_3).
\end{eqnarray}

\begin{figure} 
	\centering
	\includegraphics[width=0.9\textwidth]{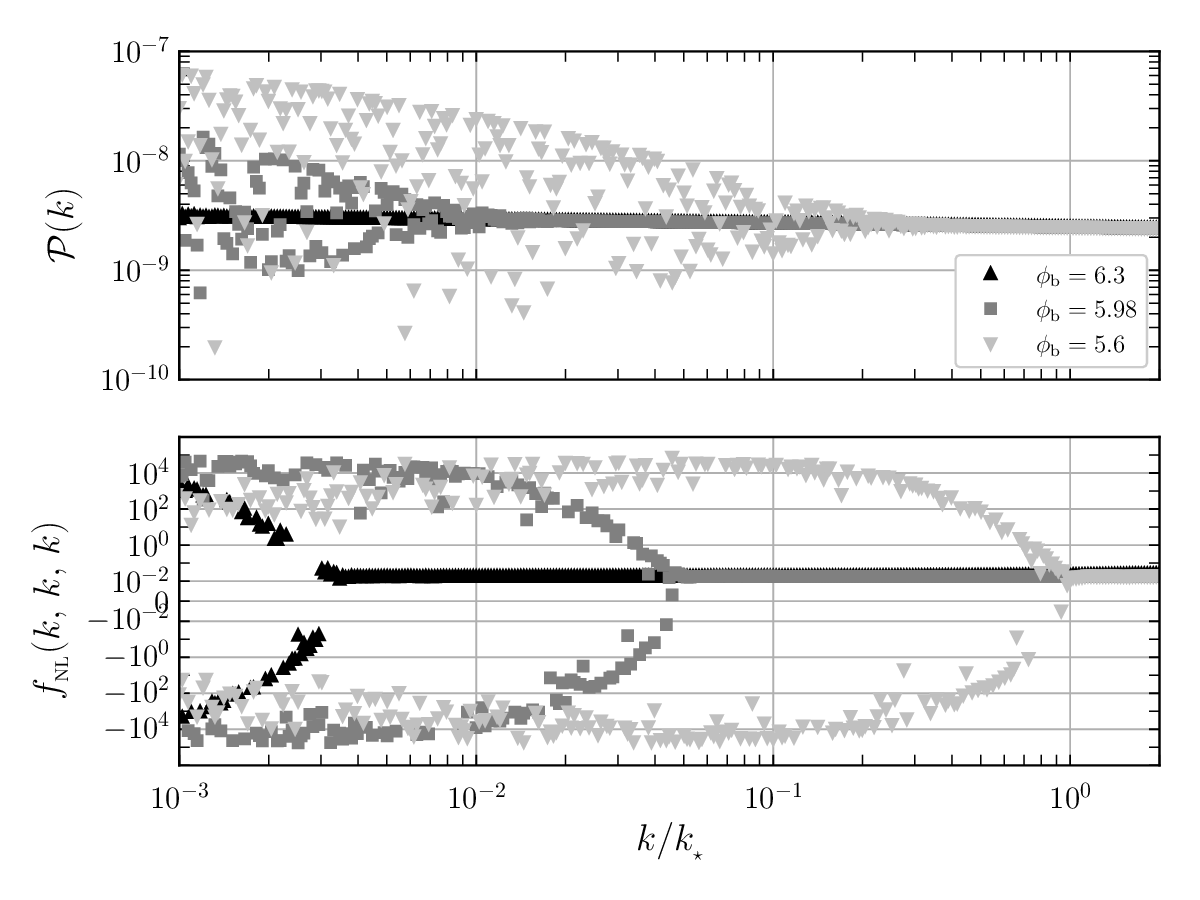}
	\caption{\label{rs:fig1} Numerical results for the primordial scalar power spectrum and $f_{_{\rm NL}}(k,\,k,\,k)$ generated in LQC as a function of $k/k_\star$ where $k_\star = 0.002\, {\rm MPc}^{-1}$ for different values of $\phi_{\rm b}$. An increase in $\phi_{\rm b}$ shifts  $k_{_{\rm LQC}}$ to more infrared scales compared to $k_\star$. In the figure $k_{_{\rm LQC}}/k_\star$ takes values $9.6\times 10^{-2}$, $4.6\times 10^{-3}$ and $3.3\times 10^{-4}$ for $\phi_{\rm b}$ equal to $5.6$, $5.98$ and $6.3$ respectively.}
\end{figure}

\subsection{Numerical Spectra}   
Both power spectrum (see, for instance, Refs.~\citenum{Agullo:2015tca,  Ashtekar:2015dja, Bonga:2015xna, deBlas:2016puz, Ashtekar:2016wpi, ElizagaNavascues:2020fai}) and bispectrum (see, for instance, Refs.~\citenum{PhysRevD.97.066021, Sreenath:2019uuo}) generated in LQC have been studied thoroughly. Here we brief the essential aspects of the numerical calculation and summarise the results. Due to the test field approximation, computation of power spectra and bispectra can be done in two stages. In first, one computes the background quantities and in second evolve the perturbations to this background and compute the correlations. 
\par 
We have worked with a simple quadratic potential to govern the scalar field $\phi$. The mass of the scalar field is set to be $6.39\,\times 10^{-6} {\rm M_{\rm Pl}}$ so that the amplitude of power spectrum at shorter wavelengths is consistent with the observation by Planck.  We set the supremum value of energy density ($\rho_{\rm sup}$) at the bounce to be  $258.9 \,{\rm M_{\rm Pl}}^4$. We impose zeroth order adiabatic initial conditions for primordial perturbations at a time much before the bounce when the observable modes were adiabatic and evolved it through the bounce and till a time when all modes of interest become super-Hubble. 
We have performed these calculations using \verb|class_lqc|\cite{PhysRevD.97.066021, Sreenath:2019uuo}. 
\par
Simulations show that universe undergoes a bounce and slow-roll inflation sets in soon after. Since perturbations evolve both through the bounce and slow-roll inflation, modes with longer wavelength get excited during bounce as well as during slow-roll regime leading to a deviation from nearly scale-invariant behaviour as expected in standard slow-roll inflation.
Numerically evaluated power spectra and non-Gaussianity function in the equilateral limit ($f_{_{\rm NL}}(k,\,k,\,k)$) for different values of scalar field at the bounce, $\phi_{\rm b}$, is shown in \fref{rs:fig1}.  Plots illustrate the following features of the spectra: (i) Bounce introduces a new scale in the problem, $k_{_{\rm LQC}}$. Studies show that this scale is related to the curvature at the time of bounce, (see, for instance, figure 4 of Ref. \citenum{K:2023gsi}) $R_{\rm b}$, as $k_{_{\rm LQC}}\, \equiv a(\eta_{\rm b})\sqrt{R_{\rm b}/6}\,\approx\,a(\eta_{\rm b})\sqrt{\kappa\rho_{\rm sup}}$. (ii) Modes $k >> k_{_{\rm LQC}}$ are not affected by the bounce and hence their power spectrum and bispectrum are nearly scale-invariant as in slow-roll inflation. (iii) Modes $k\lesssim k_{_{\rm LQC}}$ are excited during the bounce and hence depart from scale invariance. (iv) Power spectra of modes $k \lesssim k_{_{\rm LQC}}$ are scale-dependent and amplified. (v) Non-Gaussianity function of modes $k\lesssim k_{_{\rm LQC}}$ is oscillatory and amplified. (vi) Non-Gaussianity function becomes scale-dependent at a larger wavenumber than the power spectrum. In this sense, the scalar bispectrum is more {\it sensitive} to the bounce than the power spectrum. (vii) Effect of change of $\phi_{\rm b}$ is a shift in the spectra with respect to observable wavenumbers. 
The last point can be understood as follows. The value of $\phi_{\rm b}$ fixes the amount of expansion between the bounce and the epoch at which, say, the pivot scale leaves the horizon.  Higher the $\phi_{\rm b}$, larger the amount of pre-inflationary expansion. A larger amount of pre-inflationary expansion will make $k_{_{\rm LQC}}$ more infrared with respect to the pivot scale. Hence, increasing the value of $\phi_{\rm b}$ will shift the scale-dependent features of LQC to infrared scales making them not observable. 

\subsection{Compatibility with Planck}
Planck has observed a CMB that is consistent with a Gaussian and nearly scale-invariant primordial perturbations. In the light of this it is pertinent to ask whether primordial perturbations generated in LQC is consistent with these observations. Planck has found that the temperature power spectrum is consistent with a nearly scale-invariant primordial spectrum with very less error for multipoles greater than $\ell \approx 30$. 
This implies that any departure from scale invariance such as those generated in LQC should be visible only at the lowest multipoles of $\ell < 30$. Since increasing the value of $\phi_{\rm b}$ will shift the scale-dependent features of LQC to infrared scales, observations of temperature power spectra puts a lower bound on the value of $\phi_{\rm b}$. 
The minimum value of $\phi_{\rm b}$ should be such that the departure from scale invariance should be visible only at multipoles lower than $\ell \approx 30$ (see Ref.~\citenum{Agullo:2015tca}). 
\par 
Figure \ref{rs:fig1} shows us that $f_{_{\rm NL}}$ or the bispectrum that it represents is much more {\it sensitive} to the bounce than the power spectrum, {\it i.e.} bispectrum deviates from scale invariance at a higher wavenumber than the power spectrum. Further, Planck's measurement finds the value of $f_{_{\rm NL}}$ to be consistent with zero. This seems to indicate that for LQC to be consistent with observations, we need to work with an even larger value of $\phi_{\rm b}$ than demanded by constraints from the temperature power spectrum. This is a valid, but non-interesting, way in which LQC would be compatible with observations. 
It is non-interesting since a very large $\phi_{\rm b}$ would imply that no signatures of the bounce will be visible in our Universe. In such a scenario, we will not be able to verify LQC with cosmological observations.
\par 
However it turns out that CMB bispectrum does not impose additional constraints on $\phi_{\rm b}$. Due to the highly oscillatory nature of the primordial bispectrum, it does not leave any imprints in the observed CMB bispectrum. We will discuss this in the next section. See Ref.~\citenum{K:2023gsi} for a more detailed account. 
\section{CMB reduced bispectra generated in LQC\label{rs:sec3}}
Signals of primordial non-Gaussianity can in principle be observed in higher-order correlations of CMB temperature anisotropies and polarisation (see, for instance, Refs.~\citenum{Durrer:2008eom, Dodelson:2003ft, Weinberg:2008zzc, PhysRevD.63.063002, PhysRevD.76.083523, Liguori:2010hx, Fergusson:2010dm}). If primordial scalar three-point function is non-zero, it will leave its imprints in the three-point function of temperature fluctuations and electric polarisation. The three-point functions of their multipole coefficients 
are related to the primordial three-point function of curvature perturbations by 
\begin{eqnarray}\label{rs:eq11}
\langle a_{_{{\ell_1} {m_1}}}^{\text{\tiny X}}\, a_{_{{\ell_2} {m_2}}}^{\text{\tiny Y}}\, a_{_{{\ell_3} {m_3}}}^{\text{\tiny Z}}\,\rangle\,& =&\, \left( 4\pi\right)^3\, \left(- i\right)^{\ell_1+\ell_2+\ell_3}\, \int \frac{d^3 k_1}{(2\pi)^3} \int \frac{d^3 k_2}{(2\pi)^3} \int \frac{d^3 k_3}{(2\pi)^3} \Delta_{_{\ell_1}}^{\text{\tiny X}} \Delta_{_{\ell_2}}^{\text{\tiny Y}} \Delta_{_{\ell_3}}^{\text{\tiny Z}}\, \nonumber\\
&\times&\langle {\cal R}_{k_1} {\cal R}_{k_2} {\cal R}_{k_3}\rangle\, Y_{_{\ell_1 m_1}} (\hat k_1)\, Y_{_{\ell_2 m_2}} (\hat k_2)\, Y_{_{\ell_3 m_3}} (\hat k_3),
\end{eqnarray}
where ${\rm X}$, ${\rm Y}$, ${\rm Z}$ indicate either temperature fluctuation (T) or electric polarisation (E), $a^{\text{\tiny X}}_{\ell\,m}$ are their multipole coefficients, $\Delta^{\text{\tiny X}}_{_\ell}$ is the corresponding transfer functions and $Y_{\ell m}$ are spherical harmonics.  For isotropic theories, we focus on a quantity known as the CMB reduced bispectrum ($b_{\ell_1,\ell_2,\ell_3}^{\text{\tiny XYZ}}$) which is defined as follows, 
\begin{eqnarray}\label{rs:eq12}
b_{\,\ell_1\,\ell_2\,\ell_3}^{\text{\tiny XYZ}} \, &=&\, \left(\frac{2}{\pi}\right)^3\, \int x^2 d x\, \int d k_1\, \int d k_2\, \int d k_3\, \left( k_1\,k_2\,k_3\right)^2\, B_{\mathcal{R}}(k_1,k_2,k_3)\,\nonumber\\
&\times&\, \Delta_{_{\ell_1}}^{\text{\tiny X}} \Delta_{_{\ell_2}}^{\text{\tiny Y}} \Delta_{_{\ell_3}}^{\text{\tiny Z}}\, j_{\ell_1} (k_1\,x)\, j_{\ell_2} (k_2\,x)\, j_{\ell_3} (k_3\,x).
\end{eqnarray}
\par 
To study the imprints of non-Gaussianity generated in LQC on the CMB bispectrum, and hence to check the viability of LQC, we need to evaluate the reduced bispectrum. As is clear from \eref{rs:eq12}, in order to evaluate $b_{\ell_1\ell_2\ell_3}^{\text{\tiny XYZ}}$, we need to perform a computationally intensive 4-dimensional integral. This integral involves an integration over variable $x$ and three over wavenumbers of the primordial bispectrum combined with the product of six highly oscillatory functions. This computation can be simplified if we use an analytical expression for the primordial bispectrum. Such an analytical expression will save time required for numerically computing it.
\par 
For $f_{_{\rm NL}}(k_1,k_2,k_3)$ generated in LQC, see \fref{rs:fig1}, we observe that modes with wavenumbers much larger than $k_{_{\rm LQC}}$ are nearly scale-invariant and the bispectrum is similar to that generated in slow-roll inflation. Further, at scales comparable to or smaller than $k_{_{\rm LQC}}$, $f_{_{\rm NL}}(k_1,k_2,k_3)$ is amplified and oscillatory about the slow-roll like spectrum. Thus, we can model the bispectrum generated in LQC as a sum of a nearly scale-invariant bispectrum similar to that generated in slow-roll and a part which captures the amplified and oscillatory spectra generated due to the bounce. 
We model the shape of the nearly scale-invariant bispectrum  using the local template (see Ref.~\citenum{PhysRevD.63.063002}) 
\begin{eqnarray}\label{rs:eq13}
	B_{\cal R}^{\,\rm local}(k_1, k_2, k_3)\, &=&\, - \, \, \frac{6}{5}\,(2\pi^2)^2\,{\mathfrak f}^{\rm \,local}_{_{\rm NL}}\,\biggl(\, \frac{{\cal \widetilde P_R}(k_1)}{k_1^3}\,\frac{{\cal \widetilde P_R}(k_2)}{k_2^3}\, 
	+\, \frac{{\cal \widetilde P_R}(k_2)}{k_2^3}\,\frac{{\cal \widetilde P_R}(k_3)}{k_3^3}\, \nonumber\\
	&& +\, \frac{{\cal \widetilde P_R}(k_3)}{k_3^3}\,\frac{{\cal \widetilde P_R}(k_1)}{k_1^3} \biggr),\nonumber\\
\end{eqnarray}
where ${\cal \widetilde P_R}(k_1)\, =\, A_s (k/k_\star)^{n_s-1}$. To be consistent with the behaviour of $f_{_{\rm NL}}(k_1,k_2,k_3)$ at large wavenumbers, we fix ${\mathfrak f}^{\rm \,local}_{_{\rm NL}}\, =\, 10^{-2}$. Note that this amplitude of ${\mathfrak f}_{_{\rm NL}}$ is of the same order of magnitude as that predicted in slow-roll inflation. In order to model the shape of the bispectrum for modes $k \lesssim k_{_{\rm LQC}}$ which are highly oscillatory and scale-dependent, we approximate the modes to be of the form ${\rm e}^{-i\,k\,\eta}$ (see Refs.~\citenum{PhysRevD.97.066021, K:2023gsi, agullo2021large}). We can then evaluate the integral involved in \eref{rs:eq5}, under certain approximation,  to obtain the shape of bispectrum generated due to bounce as
\begin{eqnarray}\label{rs:eq14}
	B_{\cal R}^{\,\rm bounce}(k_1, k_2, k_3)\, &=&\,-\,\frac{6}{5}\,(2\pi^2)^2\,{\mathfrak f}^{\rm \,bounce}_{_{\rm NL}}\, \biggl( \frac{{\cal P_R}(k_1)}{k_1^3}\,\frac{{\cal P_R}(k_2)}{k_2^3}\,\frac{{\cal P_R}(k_3)}{k_3^3}\biggr)^{1/2}
	\nonumber\\
	&&\times\,{\rm e}^{-0.647\,\frac{k_1\,+\,k_2\,+\,k_3}{k_{_{\rm LQC}}}} \,\sin\left( \frac{k_1\,+\,k_2\,+\,k_3}{k_I} \right) .
\end{eqnarray}
In the above, we set ${\mathfrak f}^{\rm \,bounce}_{_{\rm NL}}\, =\, 1\,{\rm M_{Pl}}^{-3/2}$. Further, we model the power spectrum generated in LQC as
\begin{equation}\label{rs:eq15}
	\mathcal{P}_{\cal R}(k)=A_s\begin{cases}
		(\frac{k}{k_{_{\rm I}}})^{6}(\frac {k_{_{\rm I}}}{k_{_{\rm LQC}}})^{-0.6}\,\,\,\,\, if \,\,\,\, {k} \leq {k_{_{\rm I}}},\\
		(\frac{k}{k_{_{\rm LQC}}})^{-0.6} \,\,\,\,\,\, if \,\,\,\,{k_{_{\rm I}}}<{k}\leq {k_{_{\rm LQC}}},\\
		(\frac{k}{k_{_{\rm LQC}}})^{(n_s-1)}\,\,\,\, if\,\,\,\, {k}>{k_{_{\rm LQC}}},
	\end{cases}
\end{equation}
where $k_{_{\rm I}}$ is the scale set by the value of curvature at the onset of inflation.
Thus, the complete template for the bispectrum of curvature perturbation generated in LQC is the sum of these two templates;
\begin{eqnarray}\label{rs:eq16}
	B_{\cal R}^{\,\rm LQC}(k_1, k_2, k_3)\,=\, B_{\cal R}^{\,\rm local}(k_1, k_2, k_3)\,+\, B_{\cal R}^{\,\rm bounce}(k_1, k_2, k_3)\,.
\end{eqnarray}
\begin{figure} 
	\centering
	\includegraphics[width=0.9\textwidth]{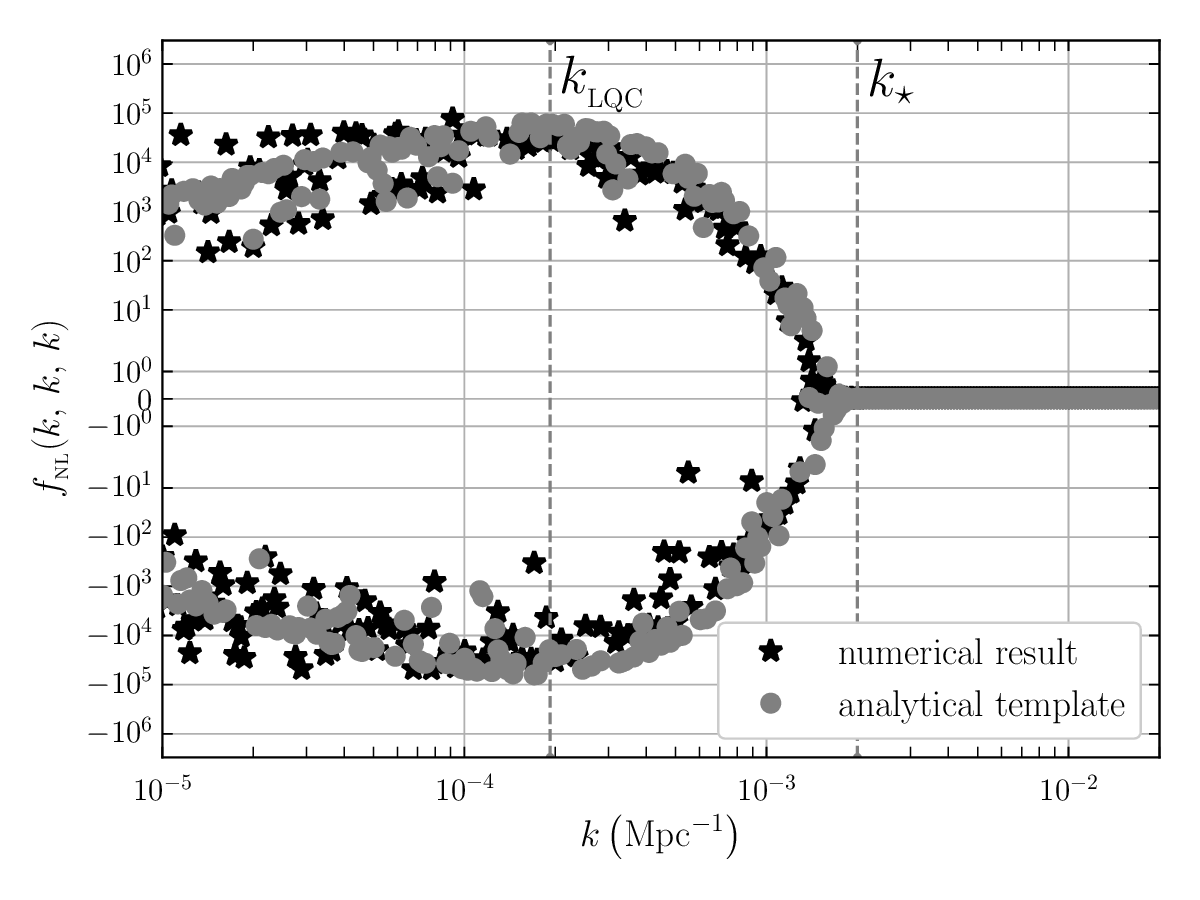}
	\caption{\label{rs:fig2} The $f_{_{\rm NL}}(k,\,k,\,k)$ generated in LQC obtained numerically (black stars) and using the analytical template (grey dots). 
	}
\end{figure}
We have illustrated the effectiveness of this template in \fref{rs:fig2}. As is evident, this form of analytical template qualitatively fits well with the numerical results. The above analytical template is effective not only for the equilateral limit but also for other configurations of wavenumbers, such as the folded and squeezed limits (see Ref.~\citenum{K:2023gsi} for more details). 
\par 
The analytical template presented above is in separable form. This implies that while evaluating reduced bispectrum, we can perform the integral over each wavenumber separately from the other. This property reduces the four dimensional integral to a two dimensional one: one integration over $x$ and another over $k$, further simplifying the calculation.
Using equations \eref{rs:eq13} - \eref{rs:eq16} in equation \eref{rs:eq12}, we can compute the reduced bispectrum of temperature fluctuations or electric polarisation generated in LQC.  Reduced bispectrum corresponding to the contribution from the bounce \eref{rs:eq14} is
\begin{eqnarray}\label{rs:eq17}
b_{{\ell_{1}}{\ell_{2}}{\ell_{3}}}^{{\rm bounce}}\,&=&\,
-\bigg(\frac{2}{\pi}\bigg)^{3}\frac{6}{5}\,(2\pi^{2})^{2}\,{\mathfrak f}^{\rm \,bounce}_{_{\rm NL}}\,\int_{0}^{\infty} d x\, x^2\, \biggl[A^{^\text{\tiny X}}_{\ell_{1}}(x)\,B^{^\text{\tiny Y}}_{\ell_{2}}(x)\,B^{^\text{\tiny Z}}_{\ell_{3}}(x)\nonumber\\&+&  B^{^\text{\tiny X}}_{\ell_{1}}(x)\,A^{^\text{\tiny Y}}_{\ell_{2}}(x) 
B^{^\text{\tiny Z}}_{\ell_{3}}(x)\,+\, B^{^\text{\tiny X}}_{\ell_{1}}(x)\,B^{^\text{\tiny Y}}_{\ell_{2}}(x)\,A^{^\text{\tiny Z}}_{\ell_{3}}(x)\,- A^{^\text{\tiny X}}_{\ell_{1}}(x)\,A^{^\text{\tiny Y}}_{\ell_{2}}(x)\,A^{^\text{\tiny Z}}_{\ell_{3}}(x)\biggr]\nonumber\\
\end{eqnarray}
and  the part corresponding to the contribution from the local template \eref{rs:eq13} is
\begin{eqnarray}\label{rs:eq18}
b_{{\ell_{1}}{\ell_{2}}{\ell_{3}}}^{{\rm local}} &=&\,
-\bigg(\dfrac{2}{\pi}\bigg)^{3}\, \frac{6}{5}\,(2\pi^{2})^{2}\,{\mathfrak f}^{\rm \,local}_{_{\rm NL}}\,\int_{0}^{\infty} d x\, x^2\,\biggl[E^{^\text{\tiny X}}_{\ell_{1}}(x)\,E^{^\text{\tiny Y}}_{\ell_{2}}(x)\,G^{^\text{\tiny Z}}_{\ell_{3}}(x) \nonumber\\
&+& G^{^\text{\tiny X}}_{\ell_{1}}(x)\,E^{^\text{\tiny Y}}_{\ell_{2}}(x)\,E^{^\text{\tiny Z}}_{\ell_{3}}(x) + E^{^\text{\tiny X}}_{\ell_{1}}(x)\,G^{^\text{\tiny Y}}_{\ell_{2}}(x)\,E^{^\text{\tiny Z}}_{\ell_{3}}(x)\biggl],
\end{eqnarray}
where, 
\begin{subequations}\label{rs:eq19}
\begin{align}
A^{^\text{\tiny X}}_{\ell}(x) &=\int_{0}^{\infty} dk\,\Delta_{\ell}^{\text{\tiny X}}(k)\,j_{\ell}(kx)\,\sqrt{({\cal P}_{\cal R}(k) {k}}\,\mathrm{e}^{-0.647\frac{k}{k_{_{\rm LQC}}}}\,\sin({\frac{k}{k_{_I}}}),\\
B^{^\text{\tiny X}}_{\ell}(x) &= \int_{0}^{\infty} dk\,\Delta_{\ell}^{\text{\tiny X}}(k)\,j_\ell(kx)\,\sqrt{{\cal P}_{\cal R}(k){k}}\,\mathrm{e}^{-0.647\frac{k}{k_{_{\rm LQC}}}}\,\cos({\frac{k}{k_{_I}}}),\\
E^{^\text{\tiny X}}_{\ell}(x) &= \int_{0}^{\infty} dk\,\Delta_{\ell}^{\text{\tiny X}}(k)\,j_\ell(kx)\,k^{-1}\, {\cal \widetilde P_R}(k_1),\\
G^{^\text{\tiny X}}_{\ell}(x) &= \int_{0}^{\infty} dk\,\Delta_{\ell}^{\text{\tiny X}}(k)\,j_\ell(kx)\,k^{2}.
\end{align}
\end{subequations}
The total reduced bispectra generated in LQC is 
\begin{equation}\label{rs:eq20}
b_{{\ell_{1}}{\ell_{2}}{\ell_{3}}}^{\text{\tiny{LQC}}}\, =\, b_{{\ell_{1}}{\ell_{2}}{\ell_{3}}}^{{\rm local}}\,+\, b_{{\ell_{1}}{\ell_{2}}{\ell_{3}}}^{{\rm bounce}}.
\end{equation}
\par 
\begin{figure}
	\begin{center}
		\includegraphics[width=0.9\textwidth]{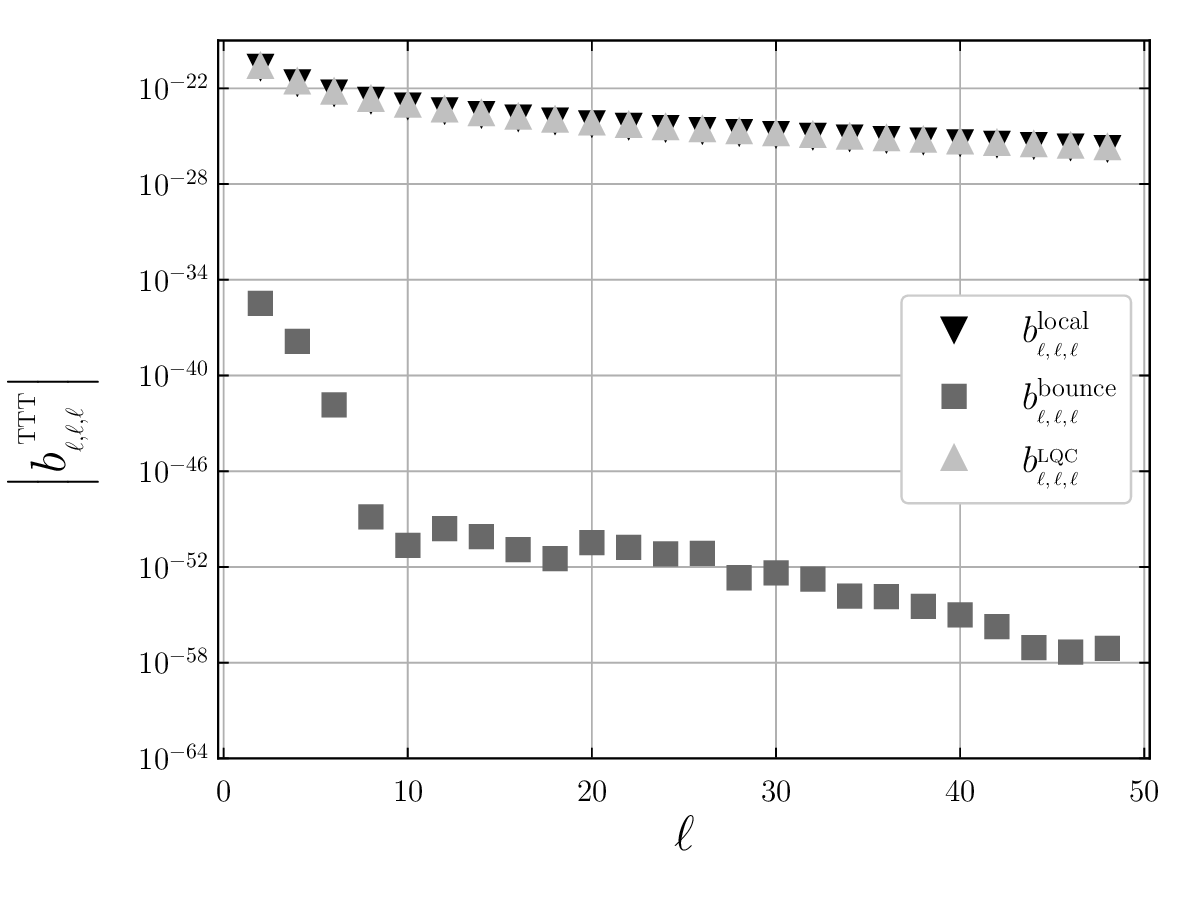}
		\caption{The contributions to the reduced bispectra of temperature correlations from the bounce, local template and their sum \eref{rs:eq20} are plotted. Since contribution from the bounce is significantly lower than the local part, the total reduced bispectrum generated in LQC is similar to that of the local template.}
		\label{rs:fig3}
	\end{center}
\end{figure}
The expressions \eref{rs:eq17} and \eref{rs:eq18} can be evaluated numerically as follows. 
First, we compute the functions \eref{rs:eq19} for different values of $\ell$ using numerical integration. These functions involve convolutions with the transfer functions ($\Delta^{\text{\tiny X}}_{_\ell}$), which can be generated using a Boltzmann code such as \verb|class|\cite{Blas:2011rf}. We used Simpson's rule with a sufficiently small step size of $10^{-8}\,{\rm MPc}^{-1}$ to numerically integrate over wavenumbers. 
After computing functions \eref{rs:eq19}, we can evaluate \eref{rs:eq17} and \eref{rs:eq18} to find the reduced bispectra. 
Integration over $x$ was carried out with a step size of 50 across the range of $x\,\in\, [0,\,40000]$. We can apply this method to evaluate the reduced bispectra $b_{{\ell_{1}}{\ell_{2}}{\ell_{3}}}^{{\text{TTT}}}$, $b_{{\ell_{1}}{\ell_{2}}{\ell_{3}}}^{\text{TTE}}$, $b_{{\ell_{1}}{\ell_{2}}{\ell_{3}}}^{\text{TEE}}$ and $b_{{\ell_{1}}{\ell_{2}}{\ell_{3}}}^{\text{EEE}}$. We show the results for $b_{{\ell_{1}}{\ell_{2}}{\ell_{3}}}^{\text{TTT}}$ in the equilateral configuration, {\it i.e.} $\ell_1\,=\, \ell_2\,=\,\ell_3\,=\,\ell$, in \fref{rs:fig3}. The results for other reduced bispectra are similar (see Ref.~\citenum{K:2023gsi}). 
\par 
In \fref{rs:fig3}, we have plotted $b_{\ell\ell\ell}^{{\rm bounce}}$, $b_{\ell\ell\ell}^{{\rm local}}$ and the total reduced bispectrum of temperature fluctuations generated in LQC, $b_{\ell\ell\ell}^{\text{\tiny{LQC}}}$. As the figure illustrates, the contribution from the bounce is significantly lower by several orders of magnitude than the contribution from the nearly scale-invariant part captured by the local template. Since the total reduced bispectrum is equal to the sum of these two contributions, we find that the reduced bispectrum of temperature fluctuations generated in LQC is similar to that generated from the local template which has been used to model a constant $f_{_{\rm NL}} \approx 10^{-2}$. This inturn implies that the bispectrum generated in LQC will be consistent with the constraints on $f_{_{\rm NL}}$ set by Planck (see, Ref.~\citenum{Planck:2019kim}). Hence LQC is viable at the level of bispectrum. Furthermore, since the amplified and oscillatory bispectrum does not leave any imprints on the CMB reduced spectrum, $b_{{\ell_{1}}{\ell_{2}}{\ell_{3}}}^{\text{TTT}}$ does not constrain the value of $\phi_{\rm b}$. 
\section{Summary and Discussion}\label{rs:sec4}
Primordial perturbations generated in LQC is scale-dependent and non-Gaussian (see \fref{rs:fig1}). The goal of this article is to discuss the constraints on LQC at the level of CMB power spectrum and more importantly bispectrum. 
\par 
Power spectrum and bispectrum have large scale-dependent amplitude at wavenumbers comparable to or smaller than $k_{_{\rm LQC}}$. In addition, bispectrum is oscillatory at these wavenumbers. At larger wavenumbers, the spectra are nearly scale-invariant as in slow-roll inflation. In LQC, $\phi_{\rm b}$ is a free parameter and its value determines the amount of expansion between the bounce and onset of inflation. Larger the value of $\phi_{\rm b}$, larger is the pre-inflationary expansion and this causes the characteristic scale $k_{_{\rm LQC}}$ to be more infrared compared to the pivot scale. The near scale invariance of power spectra of temperature fluctuations at multipoles greater than $\ell \sim 30$ allows us to constrain $\phi_{\rm b}$ (see Ref.~\citenum{Agullo:2015tca}). A lower bound on $\phi_{\rm b}$ can be set by demanding that any departure from scale invariance of the power spectrum can happen only at multipoles $\ell \lesssim 30$. Thus LQC is compatible with observations at the level of power spectrum. 
\par 
For isotropic theories the imprints of primordial bispectrum on the CMB are captured by the reduced bispectrum. Under certain assumptions, we discussed the reduced bispectrum of temperature fluctuations $b_{{\ell_{1}}{\ell_{2}}{\ell_{3}}}^{\text{TTT}}$ generated in LQC. 
The reduced bispectrum is obtained by performing a convolution of the primordial bispectrum with transfer functions and spherical Bessel functions (see \eref{rs:eq17}). It turns out that very high frequency oscillations present in \eref{rs:eq14} gets averaged out in this process and the leading contribution to the total reduced bispectrum comes from the nearly scale-invariant part of the bispectrum which is similar to that generated in slow-roll. Figure \ref{rs:fig3} illustrates this result. 
Thus the reduced bisectrum generated in LQC is very similar to that generated in slow-roll, and hence consistent with the Planck data at the level of bispectrum. This inturn implies that CMB bispectrum do not impose any constraint on $\phi_{\rm b}$. 
\par 
We shall now discuss two interesting aspects of our result. First, the perturbations generated in LQC is non-Gaussian, yet it does not leave any imprints of its non-Gaussianity on the CMB at the level of bispectrum. This imply that lack of observed CMB bispectrum does not necessarily indicate Gaussian nature. Second, the value of $\phi_{\rm b}$ is not constrained by the CMB bispectrum. This has significant implication for the possibility that CMB anomalies at low multipoles are due to the scale-dependent features of LQC (see, for instance, Refs.~\citenum{Agullo:2015aba, Zhu:2017onp, Ashtekar:2020gec, Agullo:2020wur, Agullo:2020iqv,  Ashtekar:2021izi, Agullo:2021oqk, agullo2021large, Agullo:2020fbw, Martin-Benito:2023nky}).  
Most predictions of LQC, especially those which do not take into account non-Gaussian effects, assume that the value of $\phi_{\rm b}$ is such that the scale-dependent features in the power spectrum are visible at the lowest multipoles of the CMB. If LQC had led to very large non-Gaussianity in CMB bispectra, the observations would constrain $\phi_{\rm b}$.
Since bispectrum is more {\it sensitive} than the power spectrum, it would have led to a higher value of lower bound on $\phi_{\rm b}$. A higher value of $\phi_{\rm b}$ would push the predicted features of LQC to infrared scales and hence make them difficult to observe. 
\par 
We shall now conclude the article by discussing some assumptions that we have made in making the above conclusions. First, we have assumed quadratic potential for the background. Since, in LQC, the bounce is kinetically dominated we expect the effect of physics from around the bounce to be similar for other potentials which lead to inflation as well. 
Second, in this work we have worked with zeroth order adiabatic initial conditions imposed well before the bounce for computing primordial bispectrum. 
It should be noted here that studies of some higher-order adiabatic initial conditions imposed before the bounce (see figure 9 of Ref.~\citenum{PhysRevD.97.066021}) lead to similar primordial bispectrum as discussed in \fref{rs:fig1}. 
Finally, though there are no observable signals of primordial non-Gaussianity generated in LQC in the CMB bispectrum, it may leave imprints on other higher-order correlations or on other observables. It would be interesting to test these assumptions further. 

\section*{Acknowledgments}
We would like to express our gratitude to the organizers for providing us with the opportunity to present this work. This work was supported by the Anusandhan National Research Foundation (ANRF) through Start-up Research Grant SRG/2021/001769. 

\bibliographystyle{ws-procs961x669}
\bibliography{rs}

\end{document}